\documentclass[twocolumn,prb,showpacs]{revtex4}

\usepackage{epsfig}
\usepackage{graphics}
\usepackage{graphicx}
\usepackage{dcolumn}
\usepackage{amsmath,amsfonts,amssymb,amsxtra}
\usepackage{mathrsfs}
\usepackage{bm}
\usepackage{times}

\def\cm-1{cm$^{-1}$\,}
\def\cmT-1{cm$^{-1}$/T\,}
\def\E2g{$E_{2g}$}
\def\A1g{$A_{1g}$}
\def\A1gA2g{$A_{1g}\oplus A_{2g}$\,}
\def\D3d{$D_{3d}$}
\def\2DS{$2\Delta_{S}^{E}$}

\def\2DA{$2\Delta^{A}$}
\def\D0{$2\Delta_{0}$}
\def\CaC6{CaC$_6$\,}
\def\Tc{$T_c$\,}
\def\Eg3{E$_g^{3}$}
\def\Eg2{E$_g^{2}$}
\def\Eg1{E$_g^{1}$}
\def\R-3m{$R\overline{3}m$}

\begin{document}

\title{Raman scattering from the CaC$_6$ superconductor in the presence of disorder}

\author{A. Mialitsin$^{1}$, J.\,S.~Kim$^{2}$, R.\,K.~Kremer$^{2}$ and G. Blumberg$^{1}$}

\affiliation{
$^{1}$Department of Physics and Astronomy, Rutgers University, Piscataway, New Jersey 08854-8019, USA \\\\
$^{2}$Max-Planck-Institut f\"{u}r Festk\"{o}rperforschung, 70569
Stuttgart, Germany}

\thanks{}

\date{\today}

\begin{abstract}

Polarized Raman scattering has been performed on CaC$_6$
superconductor.  We identify two of the three Raman active E$_g$
phonon modes at 440 and 1508\,cm$^{-1}$ expected for the
$R\overline{3}m$ space group of CaC$_6$. These first order
scattering modes appear along with the D and G bands around
1300\,cm$^{-1}$ and 1600\,cm$^{-1}$ that are similar in origin to
the corresponding bands in plain graphite. The intensities of the D
and G bands in CaC$_6$ correlate with degree of disorder. The D band
arises from the double resonant Raman scattering process; its
frequency shifts as a function of excitation energy with
$\sim$\,35\,cm$^{-1}$/eV. The double resonant Raman scattering
probes phonon excitations with finite wave vector
$\emph{\textbf{q}}$. We estimate the characteristic spacing of
structural defects to be on the scale of about 100\,{\AA} by
comparing the intensity of the D band and the 1508\,cm$^{-1}$ E$_g$
mode in CaC$_6$ to calibrated intensity ratio of analogous bands in
disordered graphites. A sharp superconducting coherence peak at
24\,cm$^{-1}$ is observed below $T_c$.
\end{abstract}


\maketitle

\section{Introduction}

Examples of superconductivity in graphite intercalated with alkali
metals have been known for several
decades\cite{DresselhausDresselhausAdvPh}. More recently the
occurrence of superconductivity in graphite intercalation compounds
(GICs) has been linked to the partial occupation of the inter-layer
band in those intercalated structures that display the
superconducting phase transition\cite{Csanyi, Mazin}. Experimental
studies of superconductivity in GICs were limited to very low
temperatures, since previously known GIC superconductors, like
KC$_8$ or LiC$_2$, display low superconducting transition
temperatures under ambient conditions. The synthesis of \CaC6, with
a surprisingly high superconducting transition
temperature\cite{WellerNatPh} of 11.5\,K, has spurred new research
in the area of GIC superconductivity. In addition to encouraging new
experiments, this development calls for expanding theoretical models
to explain its high \Tc. \CaC6 further merits special attention
among simple graphite derived compounds because it illustrates how
the properties of graphene sheets are altered by the change in
crystal symmetry and in electronic band structure due to interaction
with the Ca sublattice.
%D

\begin{figure}[t]
 % % Requires \usepackage{graphicx}

 %preprint 2
%\includegraphics[width=0.5\columnwidth]{CaC6str.eps}\\
%two column print
\includegraphics[width=1.0\columnwidth]{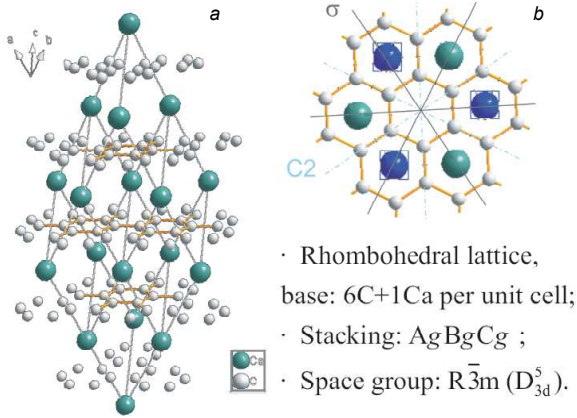}\\
\caption{(color online) Crystal structure of \CaC6. a) The layered
nature of the compound with the intercalant sandwiched between the
host graphite planes is shown in a 4 times 4 super cell. b) view on
the \CaC6 structure from the top. Thin lines indicate the two-fold
rotational axis C$_2$ and the mirror planes $\sigma$. The three-fold
rotational axis that coincides with the six-fold rotation-reflection
axis goes vertically through the paper at the point of the
intersection of the C$_2$ axes in the central hexagon. Blue colored
Ca atoms framed by squares are above the graphene layer and the
green colored ones below.
 }\label{fig:CaC6str}
\end{figure}

 The \CaC6 structure is shown in
Fig.\,\ref{fig:CaC6str}\,\emph{a}. It belongs to the space group
Nr.\,166 (\R-3m) with an $Ag Bg Cg$ stacking sequence which sets it
apart from all other known 1st stage GIC compounds
\cite{Emery1,Emery}. Here $g$ stands for the hexagonal graphene
planes stacked in a \emph{primitive fashion} (as opposed to
\emph{staggered} plane stacking of graphite). The unit cell is
rhombohedral and spans three intercalate planes. Its inversion
center lies in the middle of the graphitic hexagon in the median
plane.  Fractional contribution of Ca atoms at the corners of the
unit cell adds up to one atom per unit cell.

\CaC6 and its parent graphene structure are intuitively comparable.
Introducing Ca atoms inbetween the carbon sheets lowers the symmetry
of the D$_{6h}$ space group of graphene to the D$_{3d}^5$ space
group. This leads to a unit cell in \CaC6 with a hexagonal
cross-section area in the \emph{ab}-plane that is three times larger
than that of graphene, and a corresponding Brillouin zone (BZ) that
is three times smaller. The BZ of graphene and \CaC6 are rotated by
$30^{\circ}$ in respect to each other, thus when the graphene BZ is
folded into that of \CaC6 the equivalent K and K$'$ points fall back
onto the $\Gamma$ point. When comparing the phonon dispersion of
graphene and \CaC6 the 21 phonon branches in \CaC6 can be derived
approximately from 6 graphene phonon branches folded into the
smaller \CaC6 BZ, resulting in $3 \times 6 = 18$ branches, in
addition to the three phonon branches from the Ca sublattice.
%D

According to first principles calculations \cite{Calandra,KimPRB},
low frequency Ca vibrations and carbon out-of-plane vibrations both
contribute almost equally to most of the electron-phonon coupling
responsible for the Cooper pairs binding. Contrary to this finding,
however, based on evidence from isotope effect
measurements\cite{Hinks} it was the Ca phonons that have been found
to be primarily responsible for the mediation of superconductivity.
Lattice vibrations have also been the focus of spectroscopic
investigations: a Raman study directed at the zone center phonon
modes\cite{Hlinka} and an inelastic X-ray (IXR) scattering study
aimed at low-energy phonon branches\cite{Upton}.

The Raman study has recorded a Raman active band at 1500\,\cm-1 that
has been assigned to the high-frequency E$_g$ vibration by a rough
Raman shift coincidence with the theoretical
prediction\cite{Calandra}. Since the frequency of this band is
measured to be higher than 1447\,\cm-1 corresponding to the zone
center mode predicted by a 'frozen-phonon' density functional
calculation, it has been attempted to explain this discrepancy by
non-adiabatic behavior\cite{Saitta}. Non-adiabatic effects are
assumed to become important when a relatively short lattice
vibration period is accompanied by a relatively long electron
scattering time \cite{CastroNeto}. In such a case the Born
Oppenheimer approximation (BOA), which asserts that electrons remain
in the ground state at all times in the process of the lattice
vibration, is not applicable any more and excitations of the
electron cloud to higher states as a consequence of electron-phonon
coupling need to be considered.

 The IXR study has traced two
acoustic branches and one optical Ca sublattice branch along the
$\Gamma-$L and $\Gamma-$X directions finding good agreement with
predictions by Calandra and Mauri\cite{Calandra} barring a slight
systematic underestimation of calculated band frequencies throughout
the dispersion. The energy window of the X-ray study has been
limited to energies up to 40\,meV and in both works some of the
predicted modes have been absent in experimental data. This leaves
ample room for more in depth \CaC6 phonon studies.

Double resonant Raman scattering\cite{Martin} (DRRS) is a process
that involves light scattering mediated by phonon excitations with a
quasi-momentum greater than zero\cite{Thomsen}. DRRS allows to
access selected points in the phonon dispersion away from the
$\Gamma$ point. Raman bands associated with DRRS occur in the
presence of structural defects which are necessary to ensure
quasi-momentum conservation\cite{Thomsen}. In fact the phonon
dispersion of graphite has been investigated by means of DRRS in
considerable detail \cite{Saito}. The so-called D band, a disorder
induced feature appearing between 1300\,\cm-1 and 1400\,\cm-1 in
graphite is also seen in \CaC6 Raman spectra in the same frequency
range; however, the underlying mechanism behind DRRS in \CaC6 has
not been described. The dispersive behavior of the D
band\cite{Hlinka} in \CaC6 is a fingerprint signature of DRRS. In
the context of DRRS and disorder, Hlinka et al.\cite{Hlinka} have
highlighted the rapid sample degradation in air. It is apparent that
the interpretation of spectroscopic data is affected by sample
quality and age.

While the body of work describing superconductivity in \CaC6 is
quite substantial, there are discrepancies in the estimations of the
magnitude of the SC gap by means of tunneling spectroscopies. The
accounts vary from $2\Delta = 21.8$\,\cm-1 (Ref.\cite{Gonnelli}) to
37.1\,\cm-1 (Ref.\cite{Kurter}). The former value puts
superconductivity in \CaC6 into a weak coupling regime while the
latter raises the possibility of strong coupling. A third scanning
tunneling spectroscopy study (Ref.\cite{Bergeal}) provides an
intermediate value of 25.8\,\cm-1. This issue has been addressed in
Ref.\cite{Sanna} by proposing a distribution of gaps around the
average value of 26.3\,\cm-1 as calculated from first principles
(SCDFT). An optical investigation of reflectance of \CaC6 in far
infrared\cite{Nagel} also suggests a distribution of gaps around the
above value.

In this work we study \CaC6 crystals by polarized Raman
spectroscopy. We observe expected Raman active E$_g$ vibrational
modes at 440 and 1508\,\cm-1. The disorder induced D band appears
$\sim$\,150-200\,\cm-1 below the latter E$_g$ mode, accompanied by
weaker disorder related bands. We measure Raman spectra with
multiple excitations to highlight the dispersive nature of the D
band in the DRRS process. The prominent presence of the D band
further enables exploration of structural defects in the
investigated samples. Electronic Raman scattering reveals a sharp
superconducting pair breaking peak appearing at 24\,\cm-1 below
$T_c$. We assess how the degree of disorder correlates with the
superconducting properties.

\section{Experimental}

The high-quality \CaC6 crystals were synthesized by reacting highly
oriented pyrolytic graphite (Advanced Ceramics, Grade : ZYA) with a
molten alloy of Li and Ca at 350$^{\circ}$C for several weeks
\cite{KimPRB}. The resulting \emph{c}-oriented polycrystalline
samples were characterized by X-ray diffraction, and susceptibility
measurements. The samples for the Raman study were carefully
selected to have almost pure \CaC6 phase with a minimal amount ($< 5
\%$) of impurity LiC$_x$ phase, and to show a very sharp
superconducting transition ($\Delta T_{c}(10\%-90\%)=$ 0.1 K) with
the onset at $T_c$ = 11.4\,K. Further details on the
characterization of the samples can be found in Ref.~\cite{Emery}.
Because of the sensitivity of the lustrous sample surface to
moisture contained in air, the samples have been manipulated and
mounted in Argon atmosphere. The samples were then transferred to
and cleaved in a He-filled glove box that was enclosing the
continuous flow He-cryostat. The freshly cleaved specimen were then
immediately cooled to 5\,K and hold below 20\,K at all times for the
duration of the measurement.

To perform Raman scattering from the \emph{ab}-plane of bulk \CaC6
we have used a Kr+ laser for a range of excitation wavelengths with
2\,mW power focused to a 50 x 100 $\mu$m spot, an Oxford Instruments
cryostat for sample temperature control down to 3\,K, and a custom
triple-stage spectrometer. We have employed circularly polarized
light with the optical configurations selecting either the same or
opposite chirality. The former and the latter are respectively
referred to as Right-Right (RR) and Right-Left (RL) configurations.
The circularly polarized configurations allow to record symmetry
resolved Raman spectra. For the $D_{3d}$ point group the $E_{g}$
symmetry is selected in the RL scattering geometry and the \A1gA2g
symmetries in the RR setup \cite{Loudon}. The symmetry channels
correspond to irreducible representations of distinctive lattice
vibration modes.

\section{Raman active modes}

\begin{figure*}[t]
%preprint 4
%\includegraphics[width=1.0\columnwidth]{CaC6fig4.eps}\\
%two column print
\includegraphics[width=2.0\columnwidth]{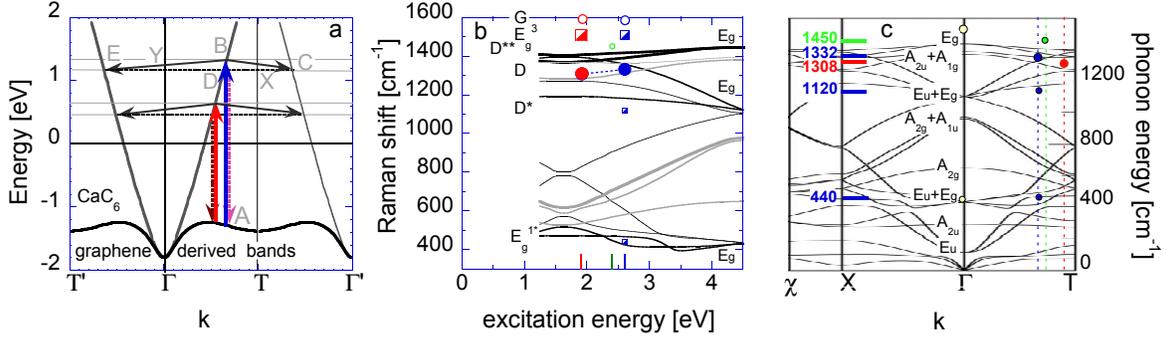}\\
\caption{(color online) Double resonance Raman scattering in \CaC6.
$a$) Scheme of the DRRS process based on graphen derived electronic
bands in \CaC6. The solid vertical arrows show the first resonant
transition. The solid diagonal arrows show the second resonant
transition in which the electron is scattered inelastically by
phonons of finite wave vector \textbf{\emph{q}}. The dashed
horizontal lines indicate defect-induced elastic recoil of the
electron. The vertical dashed lines depict the final electron-hole
recombination. Both possible scattering scenarios: inter-valley (via
T) and intra-valley (via $\Gamma$) are displayed. $b$) The solid
lines show \emph{double resonance curves}: allowed DRRS shifts a
function of excitation energy. Lines in black are derived from 'E'
branches and lines in grey from 'A' branches. DRRS curves derived
from graphene like phonon branches (E$^3_{2g}$ and A$_{2g}$) are
drawn with bold solid lines. All other (zone folded) curves are
drawn either with simple solid lines when derived from even
'\emph{g}' modes or with hair-lines for odd '\emph{u}' modes. The
observed Raman peaks are marked by symbols at the respective
excitation energies. The peak at 1450\,\cm-1 is adopted from
Ref.\,\cite{Hlinka}. $c$) The measured D, D$^{\ast}$, D$^{\ast\ast}$
and E$_g^{1\ast}$-bands positions are mapped in the \CaC6 phonon
dispersion (adopted from Ref.\cite{Calandra}) with colored and first
order Raman active E$_g$ modes with white circles.
}\label{fig:CaC6DR}
%preprint 5
%\end{figure}
%two column print
\end{figure*}

The atoms in the unit cell that remain unaffected by symmetry
operations of the space group determine the characters of the
vibrational representation $\Gamma_{vib}$. For \CaC6 we find the
respective characters to be:
\begin{table}[h]
\begin{tabular}{c|cccccc}
   D$_{3d}^{5}$
   & \quad T$_3$
   & 2C$_3$
   & 3C$_2$
   & I
   & 2S$_6$
   & 3$\sigma_d$\\
\hline

   $\Gamma_{vib}$ & \quad 21 & 0 & -3 & -3 & 0 &  1\\
\end{tabular}
\label{tab:GammaVibCaC6}
\end{table}
\newline The symmetry elements (illustrated by
Fig.\,\ref{fig:CaC6str}\,\emph{b}) are denoted as follows: T$_3$ -
translation, C$_3$ - three-fold rotation around the axis
perpendicular to the graphene plane, C$_2$ - two-fold rotation
around the axes lying in the graphene plane, I - inversion, S$_6$ -
six-fold rotation-reflection around the axis perpendicular to the
graphene plane, $\sigma_d$ - diagonal mirror planes bisecting the
angle enclosed by the C$_2$ axes. With the knowledge of the
characters table $\Gamma_{vib}$ is reduced to the direct sum
\begin{equation} \label{equ:GammaVibCaC6}
 \Gamma_{vib}={\rm A}_{1g}\oplus{\rm A}_{1u}\oplus2{\rm A}_{2g}\oplus3{\rm A}_{2u}\oplus3{\rm E}_g\oplus4{\rm E}_u \,\,.
\end{equation}
One set of A$_{2u}\oplus$E$_u$ irreducible representations
corresponds to the translational degrees of freedom (T$_z$ and
T$_{xy}$) and thus is allocated to the acoustic branches. A second
pair of A$_{2u}\oplus$~E$_u$ irreducible representations is
attributed to the mutual shift of the Ca and C sublattices
\cite{Hlinka} comprising the three lowest optical phonon branches
(see Fig.\,\ref{fig:CaC6DR}\,\emph{c}). Modes associated with Ca
movement are not Raman active.

\begin{figure}[t]
 % % Requires \usepackage{graphicx}
 \includegraphics[width=1.0\columnwidth]{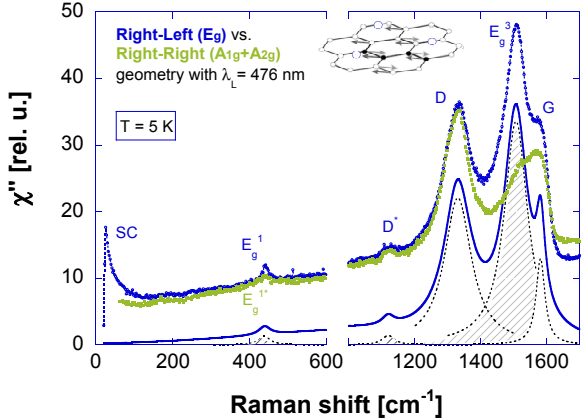}\\
\caption{(color online) Polarized Raman spectra from \CaC6 at 5\,K.
RR and RL polarizations obtained with the 476\,nm excitation.
Symbols display experimental data. Blue circles interconnected with
a dashed line correspond to the RL geometry and green circles
correspond to the RR geometry. The solid line is a fit to E$_g$ data
by superposition of five Raman oscillators on a broad continuum. The
region between 600 and 1100\,\cm-1 is featureless and has been
cropped out. }\label{fig:CaC6phon1}
\end{figure}

Polarized Raman spectra recorded at 5\,K from the \CaC6
\emph{ab}-plane are displayed in Fig.\,\ref{fig:CaC6phon1}. It shows
Raman bands observed with the 476\,nm excitation up to the Raman
shift of 1700\,\cm-1. The two juxtaposed data sets have been
obtained in different scattering configurations. The RL geometry
with circularly polarized light of opposite chirality for the
incident and scattered beams selects Raman modes of the E$_g$
symmetry. The data set collected in RL features a total of six modes
in the accessed energy range. The RR geometry with incident and
scattered light of the same chirality selects modes of the
A$_{1g}\oplus$ A$_{2g}$ symmetries. The data set collected in RR
features four bands. Depending on the presence the modes in just the
E$_g$ or in both symmetry channels we categorize the observed
features as polarized, partially polarized and depolarized.

Resolving polarization is helpful in identifying the modes. We
notice that to a large extent the RR polarization spectrum follows
the RL spectrum, with the exception of the modes at around 1100,
1500 and 1600\,\cm-1 and the low energy peak related to
superconductivity. Underlying the Raman peaks there is a broad
depolarized continuum that is phenomenologically modeled by a broad
feature peaking at around 800\,\cm-1 on top of a linear slope. We
determine frequencies of the phononic Raman bands in the RL
polarization by a fit to a superposition of Raman
oscillators\cite{LSS} and the continuum:
\begin{equation}
    \chi'' = \sum_{i} \frac{A_i \,\omega_i \,\gamma_i \,\omega}{(\omega^2-\omega_i^2)^2+(\gamma_i\,
    \omega)^2} \,\, + cont. \quad,
\end{equation}
where $A_i , \,\gamma_i$ and $\omega_i$ are the amplitude, the line
width and the Raman shift of the respective peaks. The fitted peak
parameters are listed in Table\,\ref{tab:fitEgSYMM476nm}.

\begin{table}[b]
\caption{Fit parameters for the \CaC6 Raman spectrum at $T = 5$\,K
in RL polarization excited with $\lambda_L =476$\,nm.}
\vspace*{0.1cm}
\begin{tabular}{c|c|c|c|c|c}
%\vspace*{0.2cm}
  fit parameters$\setminus$\,peaks  & E$_g^1$    & D$^{\ast}$    & D$$   & E$_g^3$   & G   \\
\hline
% & & & & &   \\
   $\omega_i$\,[1/cm] & 440    & 1120    & 1332    & 1508   & 1582   \\
%\hline
   $\gamma_i$\,[1/cm] & 36     & 44    & 100   & 86   & 32   \\
%\hline
   $A_i$\,[rel. u.$\times 10^{-3}$/cm] & 0.45    & 0.62    & 22.1   & 29.0   & 4.2   \\
\hline
  % polarized & partially     & no    & no   & yes   & partially   \\
%\hline
\end{tabular}
\label{tab:fitEgSYMM476nm}
\end{table}

 Only the 1508\,\cm-1 mode is found to be
fully polarized (present in the E$_g$ symmetry channel and entirely
absent in A$_{1g} \oplus $A$_{2g}$). This mode is identified as the
high frequency doubly degenerate carbon in-plane stretching E$_g^3$
mode (Fig.\,\ref{fig:CaC6phon1} inset). The much weaker mode at
440\,\cm-1 appears to be partially polarized. There is a sharp peak
present in the E$_g$ symmetry channel superimposed on a broader
feature appearing in both polarizations. By examining the phonon
dispersion\cite{Calandra, KimPRB}
(Fig.\,\ref{fig:CaC6DR}\,\emph{c}), we identify the additional
intensity in RL as the lowest Raman active E$_g^1$ mode which we
measure at 440\,\cm-1. The origin of depolarized modes, which we
associate with DRRS (see section IV), is different from that of
polarized modes, which are zone center modes visible due to first
order Raman scattering, and is going to be discussed in section IV.
We will discuss the graphite like G band at 1600\,\cm-1 that is
partially polarized in section VI.

The experimentally measured frequency of the E$_g^3$ mode at 5\,K
deviates from the calculated one by 4\%. \Eg1 is found to match the
frequency of 434\,\cm-1 determined from first principles
calculations \cite{Calandra} more closely with a difference of only
1.5\%. The question how and if the renormalization of zone center
modes frequencies should scale with mode frequency in the context of
first-principles calculations beyond BOA is open for discussion.

As noted above the intensities of the E$_g$ modes vary strongly. The
method of zone folding that is commonly used to qualitatively
evaluate the phonon dispersions of intercalate
compounds\cite{DresselhausDresselhausAdvPh} is helpful to discuss
this fact. The \Eg1 and the E$_g^{2}$ modes can be understood as
turned on by zone-folding of the pristine primitive graphite phonon
dispersion \cite{Dresselhaus1979}.
 In this context, the \Eg1 mode is expected to
be weak \cite{Eklund} reflecting the observation that the modulation
of the carbon layers by Ca atoms is weak. The high frequency
E$_g^{3}$ mode of \CaC6 (D$_{3d}^5$ space group) is deduced from the
$E_{2g}$ vibration in graphene\footnote{The B$_{2g}$ out of plane
graphene vibration becomes an A$_{2g}$ vibration in \CaC6. Both are
silent modes.} (D$_{6h}$) and thus is the only true not zone-folded
graphitic intra-layer Raman active mode. Accordingly the E$_g^{3}$
mode dominates the Raman spectrum in the E$_g$ symmetry channel.

The mode corresponding to E$_g^{3}$ is observed to be downshifted in
many GICs, sometimes displaying an asymmetric shape attributed to
Fano type interaction with the electronic background  (see Ref-s
\cite{Eklund, Solin} for RbC$_x$ and KC$_x$). While the frequency of
the E$_g^{3}$ phonon is clearly downshifted in \CaC6 when compared
to the parent graphite mode (see Fig.\,\ref{fig:CaC6qual} \emph{a}
and \emph{c}) we observe no asymmetry and fit the mode with a
conventional Raman oscillator.

The energy region where the E$_g^2$ mode is expected (1100\,\cm-1)
displays a depolarized feature. We believe that we observe a
disorder induced band instead of the actual zone center mode. The
Raman active A$_{1g}$ mode expected around 1380\,\cm-1 is not
observed. Our assignment of Raman bands is summarized in
Table\,\ref{tab:fitEgSYMM476nm}. We attribute two of the observed
Raman modes at 440 \cm-1 and at 1508 \cm-1 to fundamental lattice
vibrations. We assign the D and D$^{\ast}$ features to disorder
induced bands. The G band is attributed to deintercalated regions.

 \begin{figure}[t]
 % % Requires \usepackage{graphicx}
 \includegraphics[width=1.0\columnwidth]{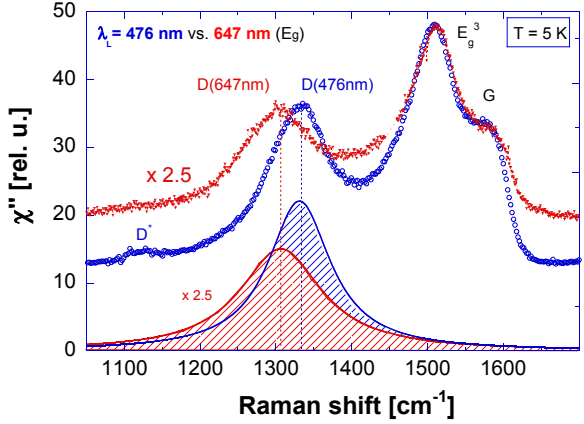}\\
\caption{(color online) \CaC6 Raman spectrum in the E$_g$ symmetry
channel excited with the 476\,nm laser line (blue circles) vs.
647\,nm (red triangles). The vertical scale of the 647\,nm
excitation spectrum has been adjusted for the E$_g^{3}$ mode to be
displayed with the same relative intensity. }\label{fig:CaC6phon2}
\end{figure}

% -------------------------------------

%preprint 3
%\begin{figure}[t]
%two column print

\section{Double Resonance Raman Scattering}

Besides the fundamental Raman frequencies, spectra of \CaC6 exhibit
features well known from disordered graphites \cite{Hlinka}. These
features are commonly labeled as D and G bands and are identified by
the rough frequency regions where they appear: around 1300\,\cm-1
for the D band and around 1600\,\cm-1 for the G band. The D band is
indicative of disorder in form of scattering centers as found in
polycrystalline samples \cite{Tuinstra}, in highly-oriented
pyrolytic graphite (HOPG) samples close to edges \cite{Cancado} and
in irradiated graphite \cite{Elman}. The G band in graphite and
graphene is attributed to the first order Raman band arising from
the E$_{2g}$ phonon, an in-plane hexagon compressing vibration. In
few-layer graphene the position of the G band depends on doping of
graphene sheets and on surface
charges\cite{YanPRL,Gupta,APSsession}. We shall discuss the origin
of the D and G band (shown in Fig-s\,\ref{fig:CaC6phon1} and
\ref{fig:CaC6phon2}) in this section.

The DRRS process couples to the disorder induced D band making
points of the phonon dispersion away from the BZ center accessible
by Raman scattering. By employing two excitation wavelengths we
measure two such points by means of the D band Raman shift. More
points in the phonon dispersion of \CaC6 will be traced with the
help of other (weaker) disorder induced Raman bands\footnote{We
observe these weak disorder modes only with the 476\,nm excitation.
These modes are apparently below the noise floor in the 647\,nm
spectra.}. Fig.\,\ref{fig:CaC6phon2}\, shows how the D band shifts
from 1308\,\cm-1 to 1332\,\cm-1 according to the excitation energy
change from 647\,nm to 476\,nm. This translates to a frequency shift
rate of 35\,\cm-1$/$\,eV in the visible light range of excitation
energies in agreement with data presented in Fig.\,1 of
Ref.\cite{Hlinka}. The observed frequency shift rate of the D band
in \CaC6 is specific to this compound and is less than the 50\,\cm-1
frequency shift rate of the D band in graphite\cite{Vidano,Matthews}
which allows to attribute the observed D band to be specifically due
to the DRRS in \CaC6, excluding the scenario that we might be
observing the graphitic D band from deintercalated regions.

The scheme for DRRS in \CaC6 (Fig.\,\ref{fig:CaC6DR}\,\emph{a})
involves graphene-derived electronic bands. There is a sequence of
four steps to this process: (1) Following the absorption of the
photon, an electron-hole pair is excited as a result of a real
inter-band transition. This event occurs at the band wave vector
\emph{k} in the BZ, where the inter-band separation energy matches
the excitation energy as indicated by a solid vertical arrow
(\emph{AB} for the blue laser line). If considering high-symmetry
directions in the \CaC6 band structure, the inter-band transition
for visible light is only possible at the wave vector
\emph{\textbf{k}} in the $\Gamma-$T direction of the rhombohedral
BZ\footnote{The Raman amplitude is going to be resonant in a region
around this point involving large BZ areas of low symmetry in the
DRRS as illustrated in Fig.\,6 of Ref.\cite{Narula}. Limiting the
discussion to the high symmetry directions of the BZ suffices for
the purpose of demonstrating the origin of the disorder induced
bands.}. This is why we chose the electronic band structure and the
phonon dispersion along $\Gamma-$T to approximate our calculation of
the DRRS effect in \CaC6. (2) Next, the electron encounters an
intra-band transition in which it is inelastically scattered by a
phonon of a finite wave vector \emph{\textbf{q}} opposite to the
direction of the solid diagonal arrow (\emph{BC}, for the blue laser
light) that illustrates the electron transition. (3) The electron is
elastically scattered back across the BZ (\emph{CD}) by a defect to
(4) recombine (\emph{DA}) with the hole in the valence band by
emission of the Raman shifted photon.

Fig.\,\ref{fig:CaC6DR}\,\emph{a} reveals two possible ways to
inelastically scatter the electron in step\,2 of the DRRS process:
(a) Across the T point into the neighboring BZ between the two
nonequivalent $\Gamma$ points, namely in the $\Gamma-{\rm
T}-\Gamma'$ direction (which is referred to as \emph{inter-valley}
scattering), or (b) across the $\Gamma$ point between the right and
left branches of the conduction band parabola inside the same BZ
(which is referred to as \emph{intra-valley} scattering). The D band
frequency is determined by finding the phonon energy in the phonon
dispersion that corresponds to the length of the phonon wave vector
$|$\emph{\textbf{q}}$|$, which is set by the second DRRS transition
by either intra-valley or inter-valley scattering.

Fig.\,\ref{fig:CaC6DR}\,\emph{b} displays allowed DRRS frequencies
as a function of excitation energy. We call this function a
\emph{double resonance curve}; there is one curve for each phonon
branch and the set of the double resonance curves is unique for any
material. The double resonance curve is obtained in two steps: (i)
We examine the \CaC6 band structure and match the laser excitation
energy to the electron wave vector \emph{\textbf{k}} where the
electron-hole pair creation can be accommodated by a corresponding
energy separation between the electronic bands. This allows us to
find the length of the phonon wave vector \emph{\textbf{q}} involved
in the inelastic scattering step of the DRRS. (ii) We read off the
phonon energy $\omega_q$ in the phonon dispersion that corresponds
to the \textbf{\emph{q}}-s obtained in the previous step for all
possible excitation energies for selected phonon branches.

The conclusion is three-fold:\newline (1) For \CaC6 there exists an
activation threshold for the disorder induced DRRS process at
1.1\,eV where the graphene derived conduction band crosses the Fermi
level. This is in difference to DRRS in graphite where the first
transition can in principle occur at arbitrary small excitation
energies as a result of inter-band transitions between the Dirac
cones in the proximity of the K points of the hexagonal BZ.
\newline
 (2) Because of the symmetry of the band structure
the inter-valley and intra-valley scattering yield the same DRRS
frequencies in the approximation that the length of the phononic
wave vector \textbf{\emph{q}} (which is the hypothenuse of the
triangles BED and BDC) is estimated by the horizontal lines DE or
DC\footnote{$q=BC=2-2k=q'=BE$, where $q$ and $q'$ are the phonon
wave vectors for inter-valley and intra-valley scattering
respectively. $q'>1$, requiring extension of the phonon dispersion
into the neighboring BZ along $|\Gamma-T|$. }.\newline
  (3) The slope of the double
  resonance curves is positive for all three graphene derived phonon
  branches (thick solid lines) in the visible light range.
Thus we expect a positive (blue) shift of the D band when the
excitation energy is increased from 647\,nm to 476\,nm. The
frequencies of observed Raman peaks are shown by symbols in
Fig.\,\ref{fig:CaC6DR}\,\emph{b}. The position of the D band for the
employed laser wavelengths is indicated by filled circles.
Considering the D band frequency range, its strong intensity and
blue-shift of the D band with excitation energy, the DRRS
characteristic displayed by the D band corresponds well to the
double resonance curve derived from the lower E$_g^3$ phonon branch.
The experimental points are below the calculated curve by
approximately 50\,\cm-1. We conclude that the D band in \CaC6
originates from the lower E$_g^3$ branch.

The D band is depolarized because phase information of the incoming
photon is lost in the process of intra-band relaxational scattering.
This is true for any DRRS induced band and thus two other
depolarized features: at 1120\,\cm-1 and at 440\,\cm-1 (labeled
D$^\ast$ and E$_g^{1\ast}$) may be interpreted as resulting from
DRRS. These modes are marked by small symbols at their respective
frequencies in Fig.\,\ref{fig:CaC6DR}\,\emph{b} and \emph{c}. We
find that E$_g^{1\ast}$ is readily associated with the upper
E$_g^{1}$ phonon branch that is mostly flat in the visible energy
region. D$^\ast$ can arise from the lower E$_g^{2}$ phonon branch.
It is 60\,\cm-1 below the calculated double resonance curve. This
error while substantial is of the same order as the error for the D
band and appears to confirm what seems to be an overestimation of
the phonon frequencies in the $\Gamma-$T direction close to the edge
of the BZ. Finally, we can treat the 1450\,\cm-1 band that has been
observed in Ref.\,\cite{Hlinka} in \CaC6 samples exposed to air the
same way. This band (labeled D$^{\ast \ast}$ and represented by a
small circle at 2.41\,eV in Fig.\,\ref{fig:CaC6DR}\,\emph{b}) can be
allocated to the highest double resonance curve originating from the
upper E$_g^{3}$ branch. Indeed, examination of Fig.\,1\,\emph{b} of
Ref.\,\cite{Hlinka} shows that the the 1450\,\cm-1 feature appears
only along a very strong D band and is absent in spectra from
freshly cleaved samples where the D band is weak. Consequently this
peak can not be attributed to a Raman active mode but is a disorder
induced DRRS band. The same argument holds also for the D$^{\ast}$
band visible in Fig.\,3 of Ref.\,\cite{Hlinka}. It confirms our
interpretation of D$^{\ast}$ as due to DRRS.

The frequencies of disorder induced bands can be plotted in the
phonon dispersion diagram yielding a measurement for points of the
phonon dispersion inside the BZ. The electronic band separation is
in resonance with the red laser excitation of $\lambda_L = 647$\,nm
(1.917\,eV) at $0.55\,|\Gamma\,{\rm T}|$ and with the blue
excitation $\lambda_L = 476.4$\,nm (2.604\,eV) at
$0.66\,|\Gamma\,{\rm T}|$, here $|\Gamma\,{\rm T}|$ is the extension
of the BZ in the respective direction. The corresponding resonant
transition pairs ($\omega_{ph}$, $q_{ph}$) are plotted for the
frequencies of the disorder bands $\omega_{ph}^{\ast}$ in
Fig.\,\ref{fig:CaC6DR}\,\emph{c}. From these experimentally measured
points of phonon dispersion we conclude that the calculation from
first principles as performed in Ref.\,\cite{Calandra}
underestimates the energy of the E$_g^3$ zone center mode, but tends
to overestimate the energy of high frequency E$_g$ branches close to
the edge of BZ in the $\Gamma-$T direction.

\section {Superconducting Gap by Electronic Raman Scattering}

\begin{figure*}[t]
%preprint 7
%\includegraphics[width=1.0\columnwidth]{CaC6SamQual.eps}\\
%two column print
\includegraphics[width=2.0\columnwidth]{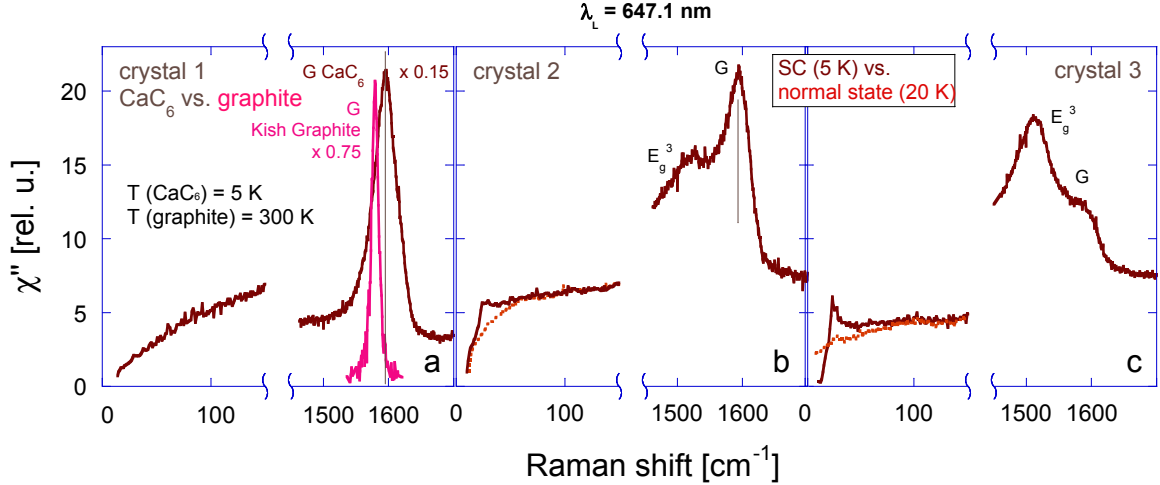}\\
\caption{(color online) The relative intensities of the E$_g^{3}$
mode (when present) and the G band in Raman spectra taken with the
647\,nm excitation in RL geometry from different samples. $a$) No SC
signature is observed in crystal 1, the E$_g^{3}$ mode is absent
also. The pink solid line shows the RL polarized spectrum from kish
graphite with its \E2g peak at 1579\,\cm-1. ($b$) The intensities of
the the \CaC6 G band and the E$_g^{3}$ mode in crystal 2 relate
approximately as 2:1. The superconducting coherence peak at
24\,\cm-1 is present but is weak and broad. c) The G band and
E$_g^{3}$ intensities in crystal 3 relate approximately as 1:2. The
strong E$_g^{3}$ mode and a sharp prominent SC coherence peak are at
the same Raman shifts as in panel (\emph{b}).}\label{fig:CaC6qual}
%preprint 8
%\end{figure}
%two column print
\end{figure*}

Below \Tc\ low energy electronic Raman scattering exhibits a sharp
SC coherence peak in the E$_g$ symmetry channel at 24\,\cm-1
(Fig.\,\ref{fig:CaC6phon1} and Fig.\,\ref{fig:CaC6qual}\,\emph{c}).
The position of the coherence peak ($\sim 3.0\, k_B T_c$) is
in-between the values $2\Delta_{ab}=21.8$\,\cm-1 reported from
directional point contact spectroscopy\cite{Gonnelli}, where '$ab$'
refers to the direction of the current flow, and 25.8\,\cm-1
determined by STM measurements\cite{Bergeal}.

The FWHM $\sim$\,12\,\cm-1 of the coherence peak obtained by
focusing the laser on a few 'high-quality' areas of the cleaved
surface is of the same magnitude as the width of the step in the
increase of reflectance below $2\Delta$ observed in Ref.\cite{Nagel}
and corresponds to the range of the anisotropic gap distribution
suggested in Ref.\cite{Sanna}. But the characteristic width of the
SC coherence peak and impurity related broadening will also
contribute to the observed line width putting limits on possible gap
distributions.

%preprint 6
%\begin{figure}[t]
%two column print

\section {The effect of intercalant disorder}
Reactivity with air humidity poses a significant challenge to
experimental investigation of \CaC6. Previous spectroscopic work on
lattice dynamics\cite{Hlinka} shows that Raman scattering is in
particular sensitive to sample aging with the characteristic E$_g^3$
mode disappearing in less than half an hour upon air exposure. X-ray
reflection study does not show comparable sensitivity \cite{Upton}.
This makes Raman scattering a preferred method to probe surface
degradation. When an imperfect spot on the sample surface is
evaluated or when the sample is exposed to humidity in air the
graphite-like D and G bands appear in the spectrum.

The disorder induced D band in graphite is associated with two main
categories of defects. First, crystallite edges or domain boundaries
\cite{Tuinstra, Cancado}, and second, defects introduced by ion
irradiation\cite{Elman} that can be understood as stray Coulomb
potentials implanted in the crystal lattice. While comparison of the
D and G band features in graphite to the D and G bands in \CaC6 can
only be qualitative due to difference in crystal structure, some
analogies to both types of disorder are observed. For
polycrystalline graphites the relative Raman intensity ratio of the
D band to the G band I(D)/I(G) has empirically been found to
increase linearly with the inverse crystallite size\cite{Tuinstra}.

Substituting I(G) of the Ref.\cite{Tuinstra} calibration with
I($E^3_g$) we find from Fig.\,\ref{fig:CaC6phon2} the ratio
I(D)/I($E^3_g$) $\approx 0.65$, which infers a hypothetical average
domain size\footnote{ With 6 C atoms in the \emph{ab}-cross section
of the \CaC6 unit cell vs. 2 atoms in the primitive cell of a
graphene layer the linear characteristic length in \CaC6 scales as
$\sqrt{3}$ in respect to that of graphite, resulting in a
$70$\,{\AA}$\times \sqrt{3} \approx 120$\,\AA\, estimate for
$L_a$(\CaC6)\,.} $L_a$(\CaC6)\, $\sim 100$\,\AA. This value is
comparable with the mean free path of $\sim 500$\,{\AA} along the
$ab$-plane estimated from the residual resistivity of $\sim
1\,\mu\Omega$cm. Additionally, this 'scale of disorder' is of the
same magnitude as the superconducting coherence length $\xi \sim
300$\,{\AA} as measured by scanning tunneling
microscopy\cite{Bergeal}. A similar I(D)/I(G) intensity ratio is
observed in annealed HOPG samples irradiated with $^{11}$B
ions\cite{Elman} at a flux rate of $5\times10^{15}$\,cm$^{-2}$.
Intercalated Ca ions when displaced from their triangular pattern in
the perfect crystalline order will have the effect of implanted
stray Coulomb potentials in \CaC6 and will display a Raman signature
that corresponds to the \emph{micro-crystallite regime} described in
Ref. \cite{Elman}. This regime that is activated for ion fluences
above $1 \times 10^{15}$ ions/cm$^{2}$ may be described as an
intermediate state between the ordered and amorphous states when
regions of disorder begin to coalesce/percolate forming islands of
ordered regions surrounded by disorder. This state is reported to be
metastable in graphite as the order can be restored by annealing.
The line width of the partially polarized G band in our \CaC6 sample
($\gamma_G = 32$\,\cm-1) also corresponds to the micro-crystallite
regime of graphite damaged by ion implantation.

In general, a sharp pair breaking peak is observed by Raman in
s-type superconductors in the clean limit\cite{Blumberg}. The
relative intensity of the SC pair breaking peak to the background
allows evaluation of the SC properties. There is a distribution of
results ranging from absence of any SC features in the Raman
spectrum (Fig.\,\ref{fig:CaC6qual}\,\emph{a}) to a sharp, well
pronounced SC coherence peak (Fig.\,\ref{fig:CaC6qual}\,\emph{c}).
Broadening and vanishing of the SC coherence peak is attributed to
different degrees of disorder caused by possible Li-ion
contamination, Ca deintercalation or aging of the cleaved surface.
The shown spectra represent the best results obtained from a set of
spots identified by a visual scan of the respective crystals
surfaces.

In Fig.\,\ref{fig:CaC6qual} we show that the relative ratio of the
E$_g^3$ mode intensity to that of the G band correlates with the
width and intensity of the SC coherence peak. We evaluate Raman
scattering in the low energy window up to 150\,\cm-1 to trace the SC
signature and in the high energy window between 1450\,\cm-1 and
1700\,\cm-1 for the phononic signature. Both measurements are
performed on the same spot of the cleaved surface and in the same
cooling cycle. Sample\,1 (Fig.\,\ref{fig:CaC6qual}\,\emph{a})
displays a smooth, close to linear slope at low energies showing no
SC peak at all. Correspondingly, at high Raman shifts the unique
E$_g^3$ phononic mode of \CaC6 is absent with only the G band
contributing to the Raman intensity. We therefore suggest that down
to the skin depth the \CaC6 structure is disturbed from its
rhombohedral form by Ca deintercalation. We compare the \CaC6 G band
to the graphitic G band recorded from a kish graphite sample to find
that in the deintercalated \CaC6 the G band is upshifted in
comparison to simple graphite. This might reflect remnant doping by
'surviving' disordered Ca atoms but should not be interpreted as a
sign of electron-phonon coupling in \CaC6 as without the E$_g^3$
mode at 1508\,\cm-1 it is not an ordered \CaC6 structure any more.
Sample~2 (Fig.\,\ref{fig:CaC6qual}\,\emph{b}) shows weak
superconductivity with a broad SC pair breaking peak on top of
underlying background intensity. In this case we do observe the
E$_g^3$ mode with about half the intensity of that of the G band.
The investigated region of sample 2 is superconducting but can be
described as \emph{low quality}. The respective shapes and relative
intensities of recorded electronic/phononic Raman modes are
characteristic of contamination. Sample~3
(Fig.\,\ref{fig:CaC6qual}\,\emph{c}) displays a sharp SC pair
breaking peak, its rhombohedral structure is intact with the E$_g^3$
mode more than two times stronger than the G band. We characterize
this sample as \emph{high quality/ low contamination} crystal. In
summary, for sample quality evaluation purpose the relative
intensity of the E$_g^3$ mode to the \CaC6 G band indicates if the
sample in question is superconducting. The E$_g^3$ mode must be
present for superconductivity to occur and the greater its intensity
relative to the G band arising from deintercalated regions the
greater portion of the sample is superconducting.

\section{Conclusion}
In conclusion we have investigated polarized Raman spectra from
\CaC6 samples with different degree of disorder. The best of the
examined samples exhibits a clear signature of superconductivity in
the form of a SC coherence peak at 24\,\cm-1. We have used this
sample to record detailed first order Raman scattering spectra. In
these spectra we have identified two fundamental E$_g$ modes and
additional graphite like D and G bands resulting from disordered and
partially non-intercalated regions. The dispersive behavior of the D
band as a function of excitation energy is a signature of double
resonant Raman scattering in \CaC6. We have calculated double
resonance curves for all phonon branches. On the basis of this
calculation we have assigned the D band to result from the lower
E$_g$ branch along the $\Gamma-$T line of the \CaC6 phonon
dispersion. By using different laser excitations we have measured
points in the phonon dispersion at finite wave vectors. From
analogies to studies of disordered graphites the investigated \CaC6
samples are best described as being in a micro-crystallite regime
with domain boundaries $\sim$\,100\,{\AA}.
% The superconducting properties of \CaC6 as seen by Raman spectroscopy
%are subject of further investigation.

We thank F.~Mauri and I.~Mazin for discussion. AM acknowledges
support by Rutgers University, the Alcatel-Lucent Foundation and the
German Academic Exchange Service.

%%%%%%%%%%%%%%%%%%%%%%%%%%%%%%%%%%%%%%%%%%%%%%%%%%%%%%%%%%%%%%%%%%

\end{document}